\documentclass[12pt]{article}

\usepackage[T1]{fontenc}

\usepackage{amsmath}
\usepackage{amssymb}
\usepackage{amsfonts}
\usepackage{graphicx}
\usepackage[dvipsnames]{xcolor}
\usepackage[colorlinks=true,urlcolor=blue,linkcolor=blue,citecolor=blue]{hyperref}
\usepackage{cite}

\usepackage[top=1.135in,bottom=1.17in,left=1.16in,right=1.16in]{geometry}
\numberwithin{equation}{section}

\allowdisplaybreaks

\newcommand{\beq}{\begin{equation}}
\newcommand{\eeq}{\end{equation}}
\newcommand{\beqa}{\begin{eqnarray}}
\newcommand{\eeqa}{\end{eqnarray}}

\newcommand{\dd}{{\rm d}}

\newcommand{\Em}{E_{\rm max}}

\begin{document}
\thispagestyle{empty}

\begin{center}

{\bf\Large   Neutrino events within muon bundles}

{\bf\Large at neutrino telescopes}

\vspace{45pt}

M.~Guti\'errez, G.~Hern\'andez-Tom\'e, J.I.~Illana, M.~Masip 

\vspace{16pt}

{\it CAFPE and Departamento de F{\'\i}sica Te\'orica y del Cosmos} \\
{\it Universidad de Granada, E-18071 Granada, Spain}

\vspace{16pt}

{\tt mgg,jillana,masip@ugr.es, ghernandezt@correo.ugr.es}

\today

\vspace{25pt}

\end{center}

\begin{abstract}
The atmospheric neutrino flux includes a component from the prompt decay of charmed hadrons
that becomes significant only at $E\ge 10$ TeV. At these energies, however, the diffuse 
flux of cosmic neutrinos discovered by IceCube seems to be larger than the 
atmospheric one.  Here we study
the possibility to detect a neutrino interaction in down-going atmospheric events at km$^3$ 
telescopes. The neutrino signal will always appear together with a muon bundle that reveals its atmospheric
origin and, generically, it implies an increase in the detector activity 
with the slant depth. We propose
a simple algorithm that could separate these events from regular muon bundles.

\end{abstract}

\newpage

\section{Introduction\label{sec1}}

The flux of atmospheric leptons, both muons and neutrinos, 
is sensitive to the multiplicity and the inelasticity
in proton-air, pion-air and gamma-air collisions, probing a forward 
kinematical region and a high energy regime that are difficult to access at colliders. 
It is apparent that an accurate description of these hadronic 
collisions is essential to connect the energy and composition of 
primary cosmic rays (CRs) with the data at neutrino 
telescopes and air-shower observatories.

One of the possibilities that has received a lot of attention throughout the years 
\cite{Volkova:1980sw,Gaisser:1988ar,Gondolo:1995fq,Costa:2000jw,Enberg:2008te,Canal:2012uh,Mascaretti:2019uqn} 
is the production of
atmospheric charm. Pions of energy above 30 GeV become less effective producing leptons in the air, 
as their decay length grows longer than their interaction length. This softens the high-energy spectrum 
of atmospheric neutrinos, changing their power law from approximately
$E^{-2.7}$ to about $E^{-3.7}$ \cite{Lipari:1993hd}. Charmed hadrons, on the other hand, 
are less frequent inside air showers, but they have a much shorter lifetime than pions and kaons. 
At energies up to the PeV scale $D$ mesons and $\Lambda_c$ baryons always
decay before they can lose energy, so their relative contribution to the atmospheric lepton flux increases
with $E$. 
It is expected that, depending on the zenith inclination,\footnote{At 
10 TeV the conventional lepton flux from light mesons is 7 times larger from near horizontal than from vertical
directions \cite{Lipari:1993hd}.} at energies around $100$ TeV  \cite{Costa:2000jw,Fedynitch:2018cbl}
this charm component may dominate the atmospheric 
lepton flux. 

Moreover, any estimate of the neutrino flux from charm decays cannot avoid a significant degree of
uncertainty. The reason is easy to understand. The primary CR flux is very steep, 
and secondary hadrons will be produced according to the
same $E^{-2.7}$ power law. Consider then an atmospheric $D$ meson of energy $E$. We may wonder
what is the most likely energy of its parent hadron. 
The $D$ may come from a hadron of energy just 10 times larger (i.e., the $D$ took
a fraction $x=0.1$ of the collision energy), but also from a parent 1000 times more energetic 
($x=10^{-3}$). Of course, a collision with
$x=0.1$ is  more unlikely than one with $x=10^{-3}$, but this may be compensated by
the fact that hadrons of energy $10^3 E$ 
are much more rare than those of just $10\,E$. It turns out that a few collisions where the charmed
hadrons take a large fraction of the collision energy could increase very substantially their production
power law in the atmosphere and thus the flux of neutrinos resulting from their decay.

Perturbative QCD calculations \cite{Cacciari:2012ny} focus 
on transverse charm and are able to reproduce 
very accurately the LHC data, but they do not include 
non-perturbative effects that may be important at forward rapidities. In particular, 
the factorization theorem used in these calculations implies that the fragmentation 
of the charm quarks produced in the collision should be
independent from the initial state. Fixed target experiments like
E791, however, contradict this scheme \cite{Alves:1996rz}. In $\pi^-$ collisions with 
Carbon and Platinum targets at 500 GeV they observed forward events of large $x$ 
where the $c$ goes into a $D^0$ or the $\bar c$ into a $D^-$
much more likely than into a $D^+$ or a $\bar D^0$, respectively. These {\it leading} charm hadrons 
appearing in the fragmentation region  share
a valence quark with the incident pion, suggesting a process of {\it coalescence} during hadronization. 
Another possibility
that may be difficult to probe at colliders is that of {\it diffractive}  charm. One may think, for
example, of a 10 TeV proton scattering off an air nucleus with a diffractive mass
 $m_p^*\approx 5$~GeV and then going into a final $ \Lambda_c \bar D$ pair that carries
all (or most of)  the initial energy. A $1\%$  component of intrinsic charm \cite{Brodsky:1980pb}
in protons and pions 
could favor these processes and imply that the forward charm 
\cite{Halzen:2016thi,Carvalho:2017zge,Goncalves:2021yvw}
contribution to the atmospheric
neutrino flux completely dominates
over the perturbative one.\footnote{For a CR
spectral index $\alpha \sim 2.7$, the flux of 
atmospheric charm is proportional to the 1.7-moment ($Z$) of the yield in hadronic 
collisions \cite{Illana:2010gh}. Therefore, $D$ mesons produced with 
a 50\% of the energy of the projectile in 0.01\% of hadronic collisions 
($Z = 3.0\times 10^{-5}$) would contribute to the prompt flux 4 times more than $D$'s 
produced with 0.1\% of the energy in 100\% of hadronic collisions ($Z=7.9\times 10^{-6}$).}

Unfortunately, the search for atmospheric neutrinos from charm decays
has been so far unsuccessful. However, IceCube observed in 2013 \cite{Aartsen:2013jdh,Abbasi:2020jmh}
a diffuse flux of cosmic neutrinos that at $E>30$ TeV is several times larger than the total 
atmospheric flux. In the 30--500 TeV region its spectral index $\alpha$ seems similar 
to what we may expect from atmospheric charm ($\alpha\approx 2.7$), 
whereas at PeV energies the cosmic flux becomes harder ($\alpha\approx 2.0$--$2.3$). 
Although this flux is a great discovery, it makes the possibility to detect 
neutrinos from charm even more difficult. In upgoing or near-horizontal events
both fluxes are indistinguishable \cite{Carceller:2017tvc}, as they are expected with the same angular 
distribution and imply a similar ratio of shower to track 
(with a muon after the interaction) events. Actually, the best fit obtained
by IceCube from the data on high energy events is no neutrinos from charm at all. Obviously, their analysis
is performed trying to minimize the atmospheric background, {\it i.e.}, cutting any events
where muons enter the detector from a down-going direction. 

Here we will explore the opposite possibility. We
will focus on down-going events, where the neutrino signal appears together with a muon bundle
that, in turn, guarantees its atmospheric origin. 
Arguably, this is what will be needed to determine the prompt neutrino flux. Other approaches 
(spectrum and lateral separation of large $p_T$ muons 
\cite{Abbasi:2012kza,IceCube:2015wro,Soldin:2018vak}) 
focus on down-going events as well.
Our analysis will involve two main aspects that we 
study by using the air shower simulator CORSIKA \cite{Heck:1998vt}: 
{\it (i)} the relation between a neutrino of given energy and the energy of
its parent air shower and {\it (ii)} the characterization of muon bundles from CR
primaries of any energy and composition (in Sections~\ref{sec2} and \ref{sec3}).
Then we will analyze the longitudinal energy depositions through 
the ice or water in a down-going event with or without a $\nu$ interaction; these depositions 
determine the detector activity at
km$^3$ observatories like IceCube or KM3NeT \cite{Adrian-Martinez:2016fdl}. Finally, we propose an algorithm
based on four observables that could be used to separate events with 
an atmospheric neutrino interaction from 
events with just stochastic energy depositions of a muon bundle (in Section~\ref{sec4}). 

\section{Neutrinos and their parent cosmic ray\label{sec2}}

Let us start with the following question. Suppose we observe an atmospheric neutrino of energy 
$E_\nu=10$ TeV entering a km$^3$ telescope from a zenith inclination $\theta_z=45^\circ$. 
What is the energy of its parent CR? Obviously, this neutrino may have been
produced by a CR of {\it any} energy $E>E_\nu$, so the actual question is: What is the 
probability distribution of the parent energy? The answer will depend
on two basic quantities: the yield of neutrinos of energy $E_\nu$ produced per proton air shower of 
energy $E$, and the primary CR spectrum and composition at $E>E_\nu$.

We may express the neutrino yield per proton shower as $f_{p \nu}(x,E)$, where $x=E_\nu/E$ is the fraction of the
shower energy taken by the neutrino. Notice that $x$ takes values between 0 and 1, that the 
integral of $f_{p \nu}(x,E)$ between these two values gives the total number of neutrinos produced inside
the shower, and that if instead we integrate $x\,f_{p \nu}(x,E)$ we will get the fraction of the shower energy carried
by all these neutrinos.

We have used CORSIKA with SIBYLL 2.3C \cite{Fedynitch:2018cbl} as the hadronic interaction model
to deduce the $\nu_e$ and $\nu_\mu$ yields from 
proton primaries of $E=(10^3$, $10^4$, $\dots$, $10^8$)~GeV, and we have obtained a simple
fit that performs well in this energy interval (see the details 
in appendix A). In Fig.~\ref{fig1} we plot the total 
yields at three different energies from $\theta_z=45^\circ$ 
(we provide the zenith angle dependence in the appendix) together with our fit.
\begin{figure}[t]
\begin{center}
\includegraphics[width=0.49\linewidth]{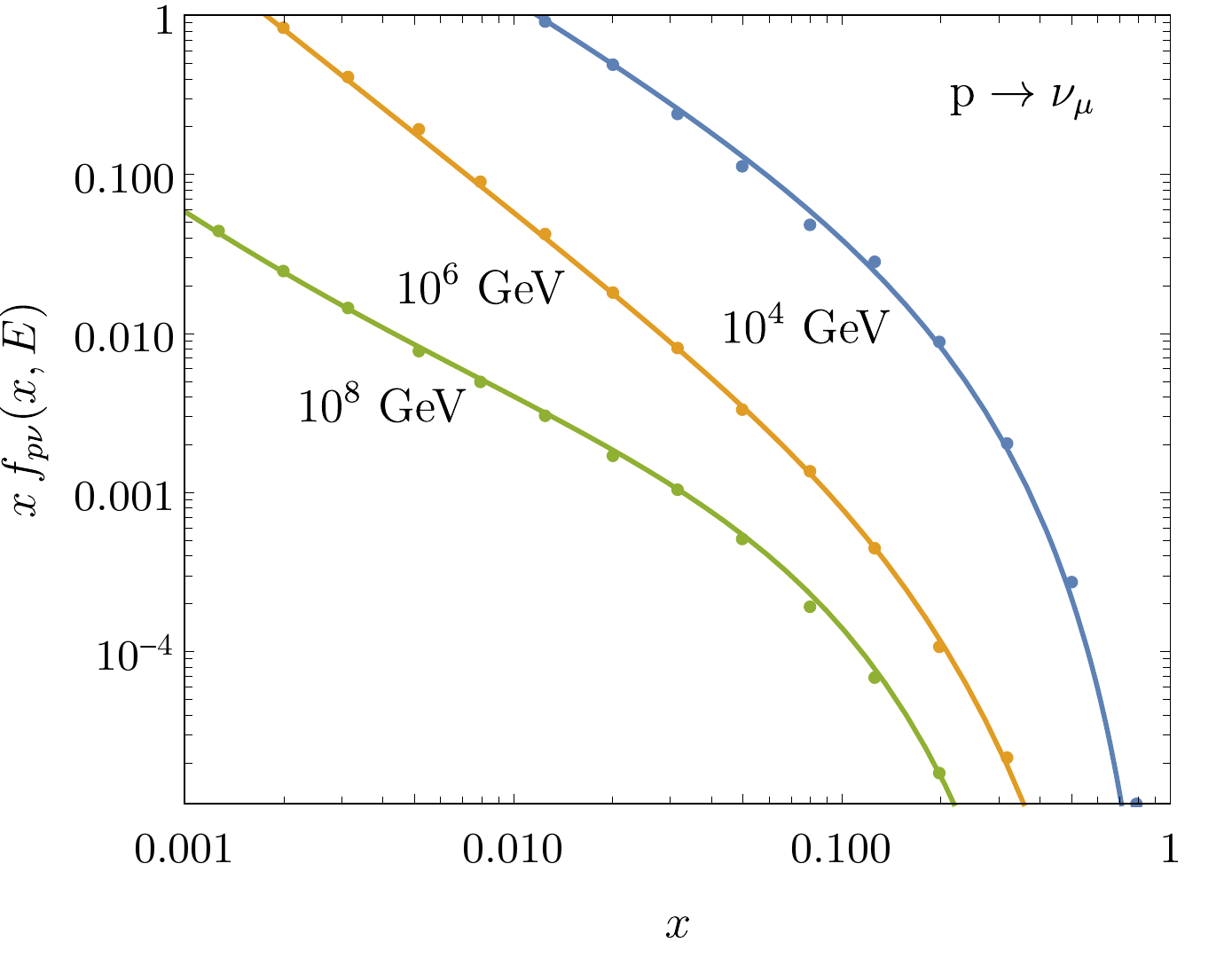} \hfill
\includegraphics[width=0.49\linewidth]{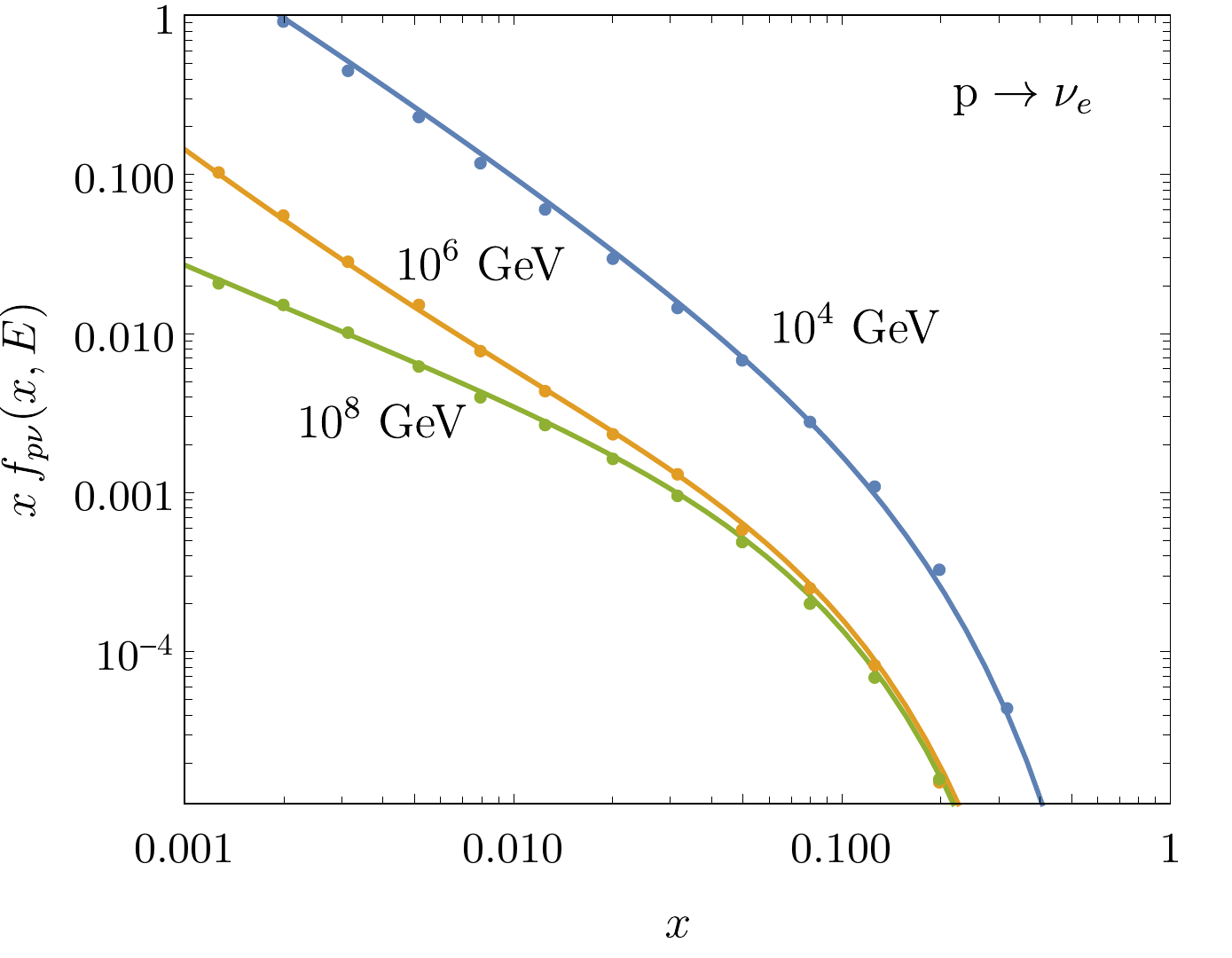} \\[2ex]
\includegraphics[width=0.49\linewidth]{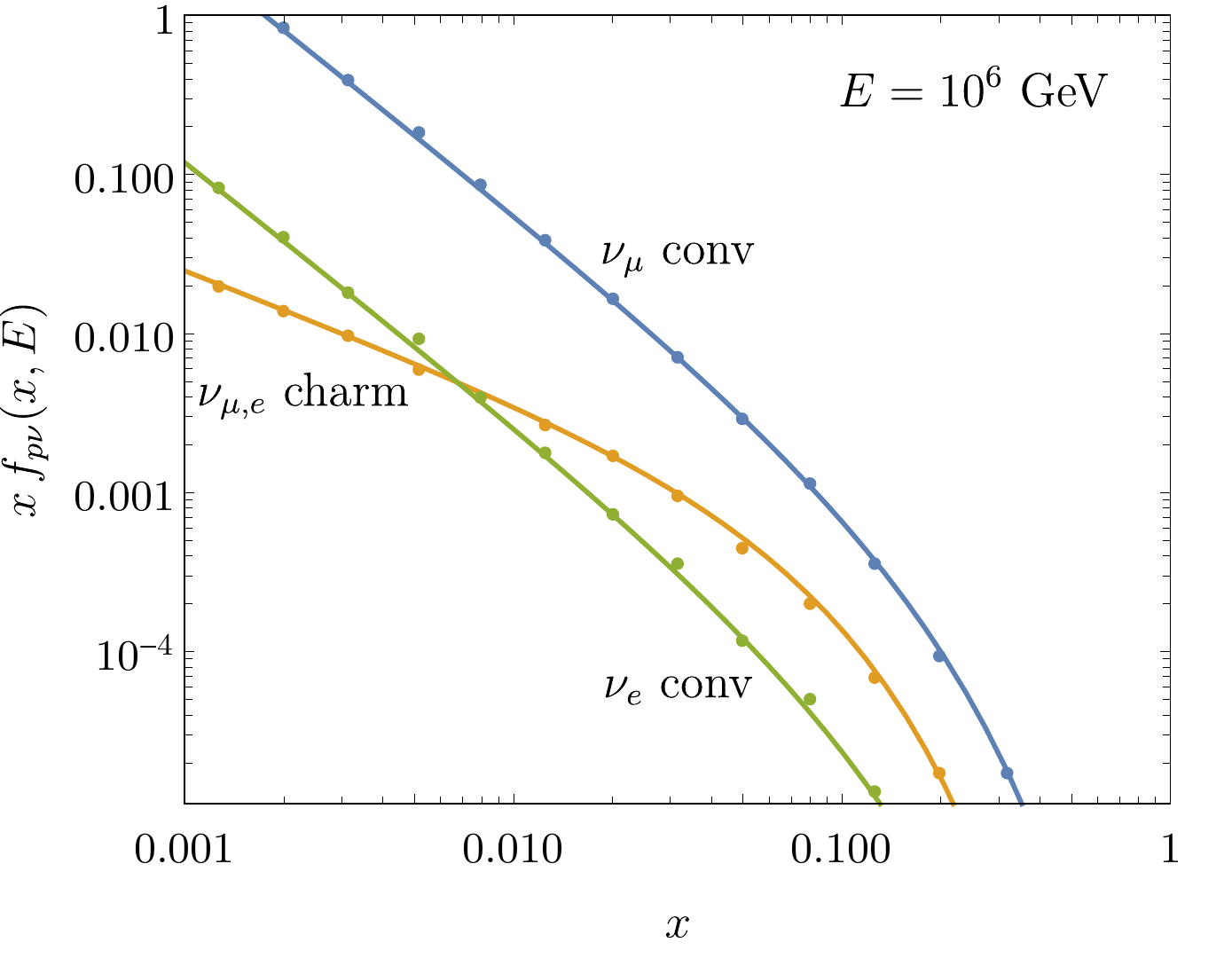}
\caption{$(\nu_\mu+\bar \nu_\mu)$ and $(\nu_e+\bar \nu_e)$ yields from 
proton showers of several energies (upper); conventional and charm components 
in both yields for a $10^6$~GeV primary (lower).
\label{fig1}}
\end{center}
\end{figure}
The plots show that lower energy showers are more likely to include a neutrino carrying a large 
fraction $x$ of the shower energy. These yields 
must be understood as the sum of two contributions: 
conventional neutrinos from pion and kaon decays plus neutrinos from the decay of charmed hadrons.
The lower plot expresses the relative contribution of these two 
components for an average $10^6$~GeV proton shower. We find that 
CORSIKA gives the same $\nu_e$ and $\nu_\mu$ yields from charm and an almost perfect scaling ({\it i.e.},
the charm contribution in this plot does not depend on $E$; we will neglect the $\approx 2\%$ $\nu_\tau$
component from $D_s$ decays). 
The plot also shows, 
for example, that in the $10^6$~GeV proton shower charm decays dominate the production of $\nu_e$'s of
$E_\nu>7$~TeV, 
or that a $\nu_\mu$ of $E_\nu > 100$~TeV inside the same shower 
is still 4 times more likely conventional than from charm.

From these yields in proton showers 
we can easily estimate the ones for other primaries, like He or Fe. In particular, 
assuming that a 
nucleus of mass number $A$ and energy $E$ is the superposition of $A$ nucleons of energy
$E/A$, we obtain
\beq
f_{A \nu} (x,E)=A^2\, f_{p\nu}(Ax,{E\over A})\,.
\eeq

As mentioned above, the second key element 
to relate an atmospheric neutrino with its parent shower is the primary CR flux. 
At energies below $E_{\rm knee}=10^{6.5}$~GeV we will assume that it is dominated by proton and
He nuclei  with slightly different spectral indices \cite{Boezio:2012rr}, 
\beq
\Phi_p = 1.3 \left( {E\over {\rm GeV}} \right)^{-2.7} {\rm particle \over GeV\,cm^2\,s\,sr}\,
;
\;\;\;
\Phi_{\rm He} = 0.54 \left( {E\over {\rm GeV}} \right)^{-2.6} {\rm particle \over GeV\,cm^2\,s\,sr}
\,.
\label{fluxpHe}
\eeq
These fluxes imply an all-nucleon flux 
$\Phi_N \approx 1.8 \,( {E/ {\rm GeV}} \,)^{-2.7}\,$${\rm [nucleon \,( GeV\,cm^2\,s\,sr)^{-1}]}$ 
and a similar number of protons and He nuclei at 
$E\approx 10$TeV.
Beyond the CR knee, up to 
$E_{\rm ankle}=10^{9.5}$~GeV, the
composition is uncertain, while the total flux becomes
$\Phi = 330 \left( {E/ {\rm GeV}} \right)^{-3.0}\,$${\rm [particle \,( GeV\,cm^2\,s\,sr)^{-1}]}$. 
Throughout our analysis we will consider the limiting
cases with a pure proton or a pure Fe composition at $E>E_{\rm knee}$ and will take a central case
where the composition is assumed to be protons and He nuclei in the proportion estimated at $E=E_{\rm knee}$.
\begin{figure}[t]
\begin{center}
\includegraphics[width=0.49\linewidth]{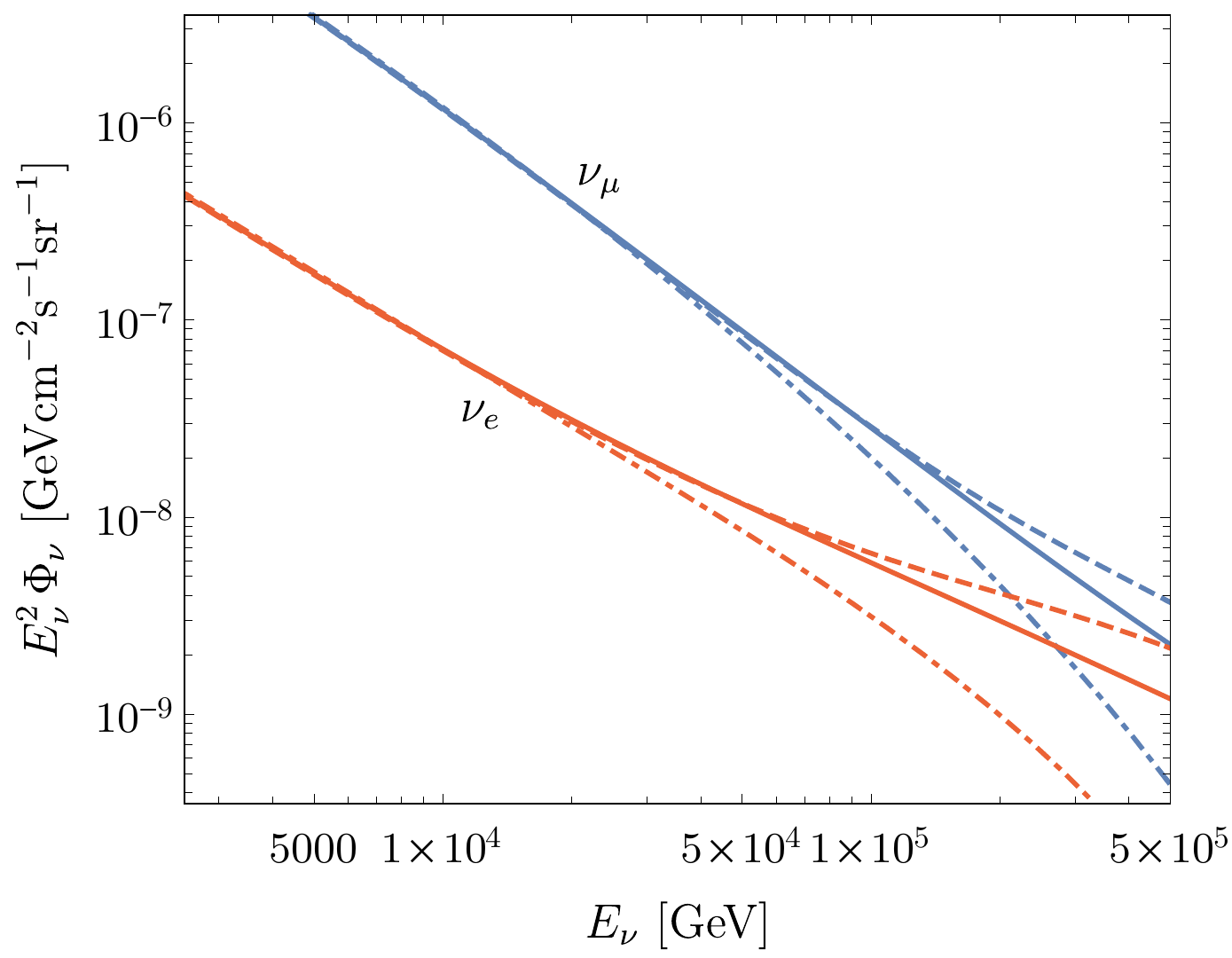}\hfill
\includegraphics[width=0.49\linewidth]{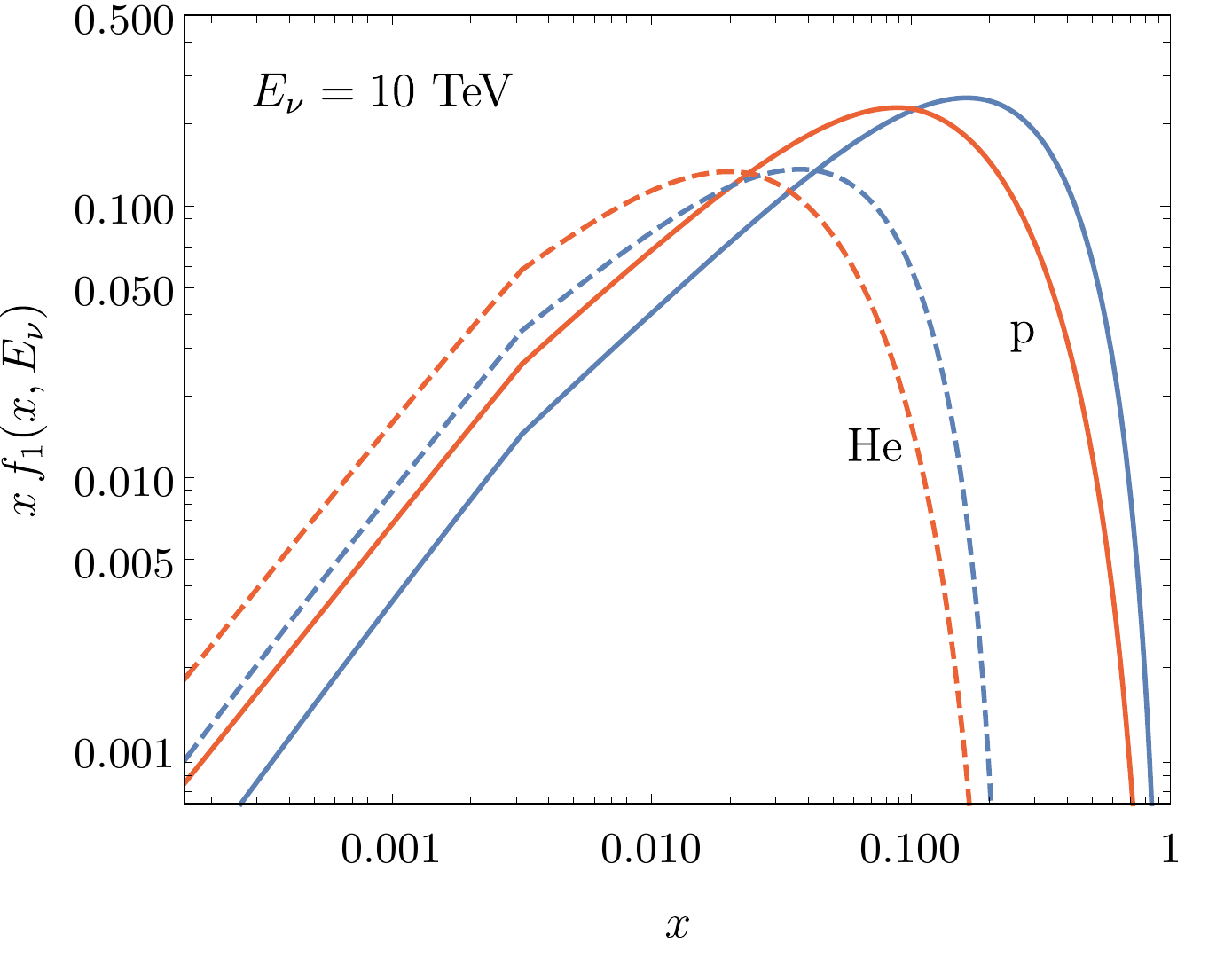}
\caption{{\bf Left.} 
Atmospheric neutrino flux for $\theta_z=45^\circ$; we include the flux for a pure proton (dashed) or
Fe (dotdashed) CR composition at $E>E_{\rm knee}$. {\bf Right.} Distribution (scaled by the relative contribution 
of each primary to $\Phi_\nu$) of $x=E_\nu/E$ at $E_\nu=10$TeV for $\nu_e$ (red) and $\nu_\mu$ (blue)
both for $p$ (solid) and He (dashed) primaries.
\label{fig2}}
\end{center}
\end{figure}

Our results are summarized in Fig.~\ref{fig2}. At 10 TeV and $\theta_z=45^\circ$ the atmospheric   
neutrino flux $\Phi_{\nu}$ includes a $6\%$ of $\nu_e$ and a $94\%$ of $\nu_\mu$. 
For a pure proton composition above $E_{\rm knee}$, $66.3\%$ ($67.5\%$) of 
$\Phi_{\nu_\mu}$ ($\Phi_{\nu_e}$) comes from proton showers, whereas this percentage decreases to 
$61.6\%$ ($61.3\%$) for a pure Fe composition. 
When the parent is a proton 
(solid lines in Fig.~\ref{fig2} Right), the fraction of energy taken by the neutrino is distributed 
according to
\beq
f_1^p(x,E_\nu)={1\over \Phi^p_{\nu}(E_\nu)}\, {f_{p\nu}(x,{E_\nu/ x})\; \Phi_p({E_\nu/ x})\over x}\,,
\eeq
where $\Phi^p_{\nu}$ is the contribution from proton primaries to $\Phi_{\nu}$. An 
analogous expression describes 
the fraction of energy taken by  neutrinos coming from a He primary
(dashed lines in the same figure). We obtain
that a $\nu_\mu$ 
(blue lines) carries in average $13\%$ of the shower energy when the primary
 is a proton or  $3\%$ when it is a He nucleus. For  electron neutrinos (red lines) 
these average fractions are a bit smaller: $9\%$ and $2\%$, respectively.

Our results
may seem somewhat surprising. It is apparent that most of the neutrinos produced by
a CR primary of energy $E$ will carry a very small fraction of the shower energy (see 
Fig.~\ref{fig1}), however, the rare events where the neutrino takes a large fraction 
of this energy dominate $\Phi_\nu$. 
The steep fall of the CR flux with the energy suppresses the contribution of 
neutrinos with a small $x$, {\it i.e.}, inside very energetic showers. 
We find, for example, that when the primary is a proton 
75\% of muon neutrinos of $E_\nu=10$TeV 
come from showers of $E<232$ TeV, and that the ratio $x=E_\nu/E$ grows even larger 
at lower neutrino energies 
({\it e.g.}, at $E_\nu=1$ TeV, $E<21$~TeV). These 
results, fully compatible with the ones 
in \cite{Fedynitch:2018cbl}, imply that most neutrino events take place 
inside relatively weak muon bundles.

\section{Leading muon and  muon bundle\label{sec3}}

Atmospheric muon neutrinos will always be produced together with a $\mu^\pm$ of similar energy.
Since these 
neutrinos carry a significant fraction of the shower energy, it follows that there will be 
a {\it leading muon} of energy well above the average muon energy in the bundle at the core of the
air shower. This leading muon will be absent in $\nu_e$ events.

It is straightforward to parametrize its energy distribution using the CORSIKA
simulations described in the previous Section. Suppose that a proton shower of energy $E$ produces a 
$\nu_\mu$ of energy $E_\nu=xE$ with $x>10^{-3}$; let us define the energy of the leading muon as 
$E_\mu\equiv e^{\alpha_\mu} x E$
({\it i.e.}, $E_\mu=E_\nu$ for $\alpha_\mu=0$). We find that the distribution of $\alpha$ can be fitted 
with\footnote{For each shower energy $E$, we just bin $x$ and find the average and the dispersion of $\alpha$
in each bin, fitting the result with this gaussian.}
\beq
f_2^p(\alpha_\mu, x, E)={1\over 0.8 \sqrt{2\pi}}\, \exp { \left( {\alpha_\mu + 2.1 +( 1.01 - 0.042 \ln E ) \ln x 
\over 0.8 \sqrt{2} }\right)^2 },
\eeq
with $E$  given in GeV. The energy distribution of the muon accompanying the neutrino produced inside
a shower of energy $E_\nu/x$ is then
\beq
\tilde f_2^p(E_\mu, x,E_\nu)={1\over E_\mu}\,f_2^p(\ln {E_\mu/ E_\nu},x, {E_\nu/ x})\,.
\eeq
This distribution will be independent from the zenith inclination of the primary but not
 its composition.
When the primary is a nucleus of mass number $A$, the distribution is
obtained just by changing $E\to E/A$ and $x\to Ax$ in the expression above. 

Suppose that a 200 TeV proton shower produces a 10 TeV $\nu_\mu$; we find that the leading
muon has in this case 
an average energy of $\langle E_\mu\rangle = 7.5$ TeV, and that with a 50\% probability 
$E_\mu<5.4$ TeV. We find remarkable that, although in average muons take more energy than
neutrinos in meson decays, the leading muon inside a shower with a very energetic 
($x>10^{-3}$) neutrino carries a smaller fraction of energy. Our result reflects that the
neutrinos emitted forward in the meson decay contribute to $\Phi_\nu$ more than the ones emitted backwards,
and in the first case the muon takes a smaller fraction of the meson energy.

\begin{figure}[t]
\begin{center}
\includegraphics[width=0.5\linewidth]{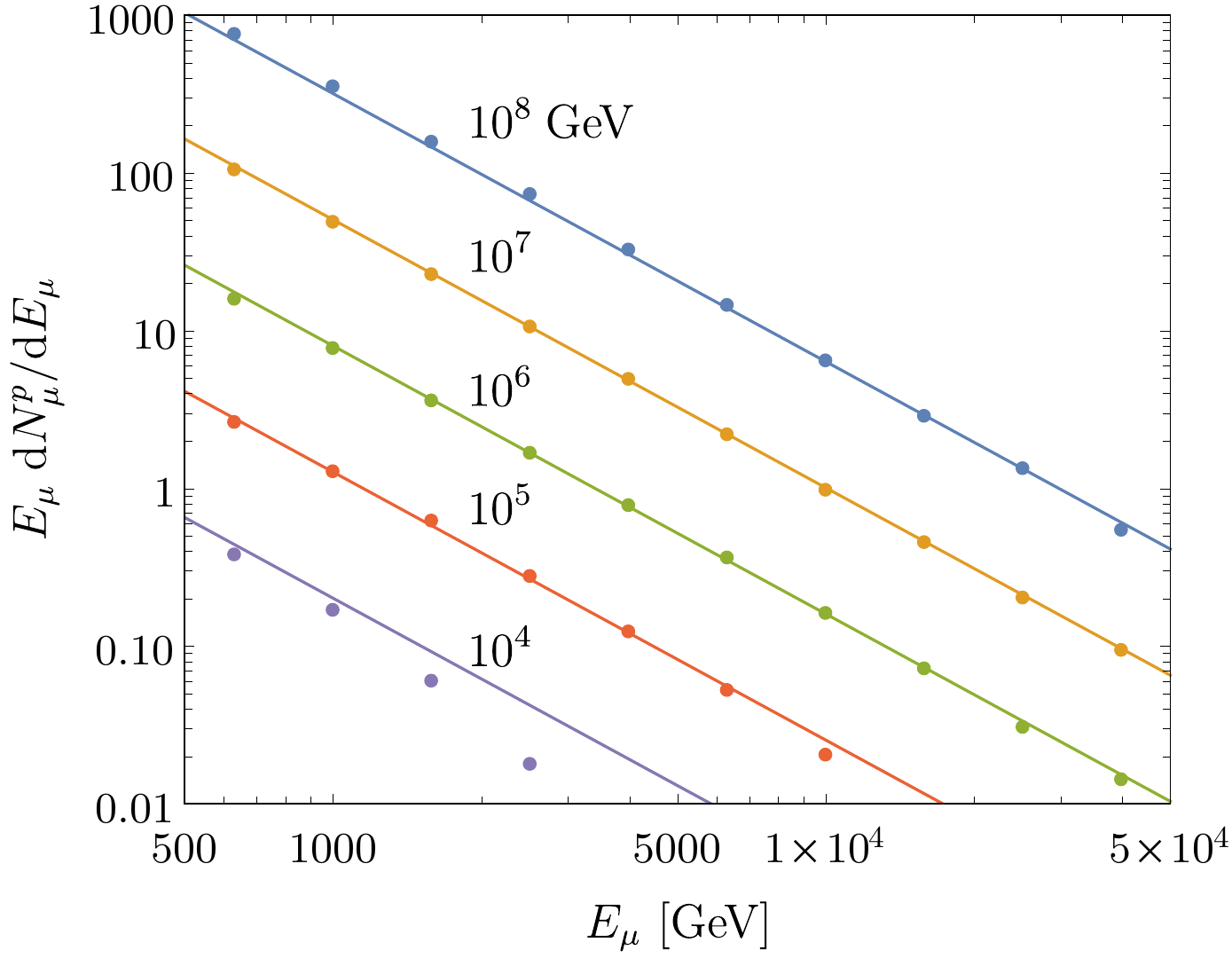}
\caption{
Spectrum of muons produced in proton showers at several energies obtained with CORSIKA ($10^4$ showers) 
together with the fit in Eq.~(\ref{muons}).
\label{fig3}}
\end{center}
\end{figure}
To study 
the possibility to detect atmospheric neutrino interactions in down-going events, a precise 
characterization of the muon bundle in the core of the air shower is also essential: we use CORSIKA to obtain 
a fit of their number and energy distribution. In a proton shower of energy $E$ from 
$\theta_z=45^\circ$ the muons of $E_\mu\ge 500$~GeV are distributed according to (see Fig.~\ref{fig3})
\beq
{\dd N^p_\mu(E_\mu,E)\over \dd E_\mu} =16\, E^{0.8}\, E_\mu^{-2.7}
\label{muons}
\eeq
(all energies in GeV) up to $E_\mu\approx 0.2 E$, with a 30\% dispersion with respect to this central value. 
The total number of muons $N_\mu^p(E)$ 
is then obtained by integrating this expression. 
Again, we will approximate
the bundle in the shower started by a nucleus as the  sum of $A$ proton bundles of energy $E/A$. As for 
the zenith
angle dependence, it can be approximated by the same factor that multiplies the conventional yield 
in the expression (\ref{theta}) given in the appendix. Our results on the spectrum and number of muons in
a bundle are consistent with the ones discussed in \cite{IceCube:2015wro}.

Once the muons penetrate the ice or water, they will lose energy  through four basic processes: ionization,
pair production, bremsstrahlung and photohadronic interactions. We will use the differential cross sections
$\dd \sigma/\dd \nu$ for these processes in \cite{Groom:2001kq}, where  
$\nu$ is the fraction of the muon energy deposited in these collisions with Hydrogen and Oxygen nuclei.
To simulate the propagation of each individual 
muon we  define steps of 25 m and  separate soft
collisions that imply a continuous energy loss from harder stochastic processes. In the first type we include
both ionization and radiative collisions of $\nu<10^{-2.5}$.

\begin{figure}[t]
\begin{center}
\raisebox{-0.5\height}{\includegraphics[scale=0.52]{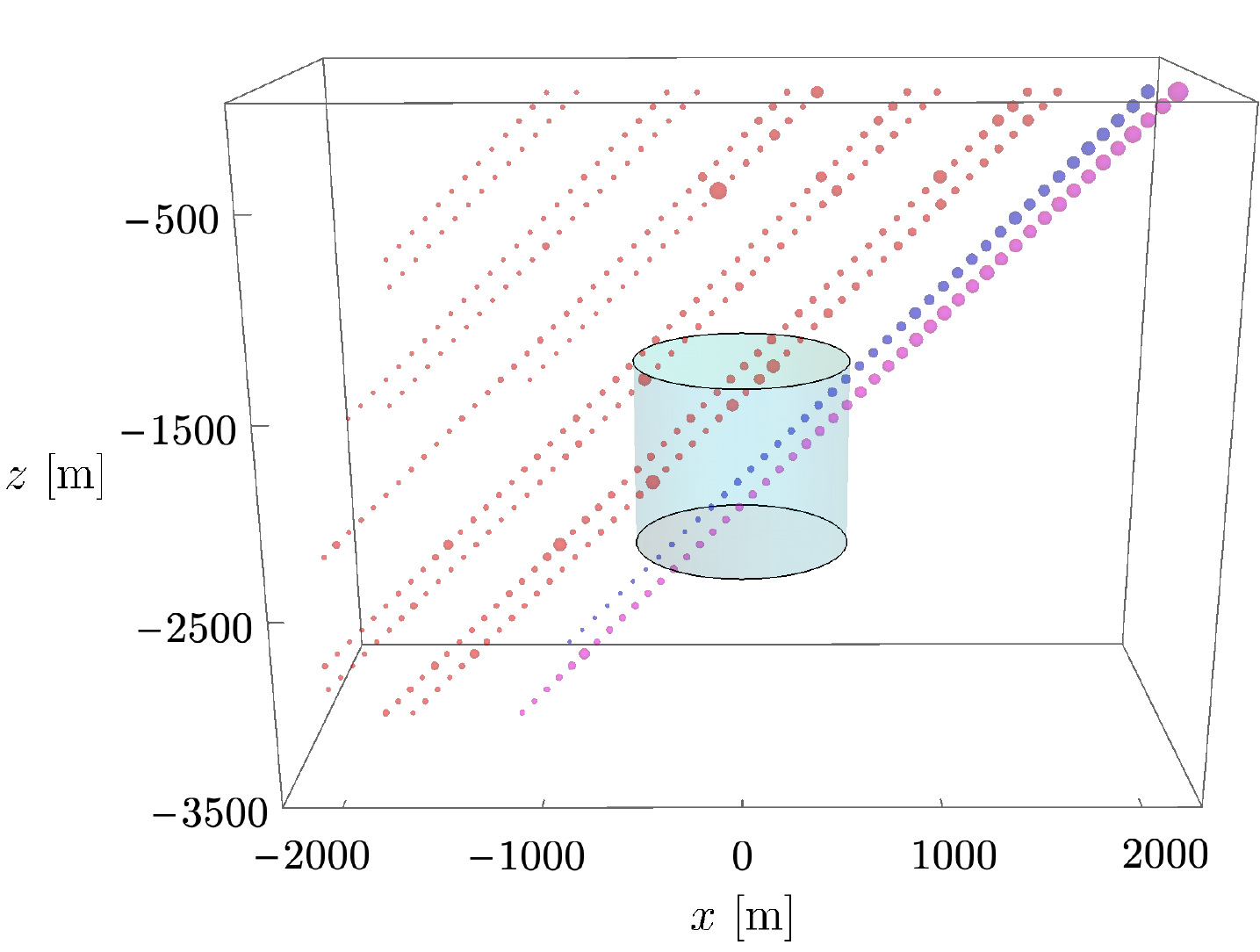}}\hfill
\raisebox{-0.5\height}{\includegraphics[scale=0.52]{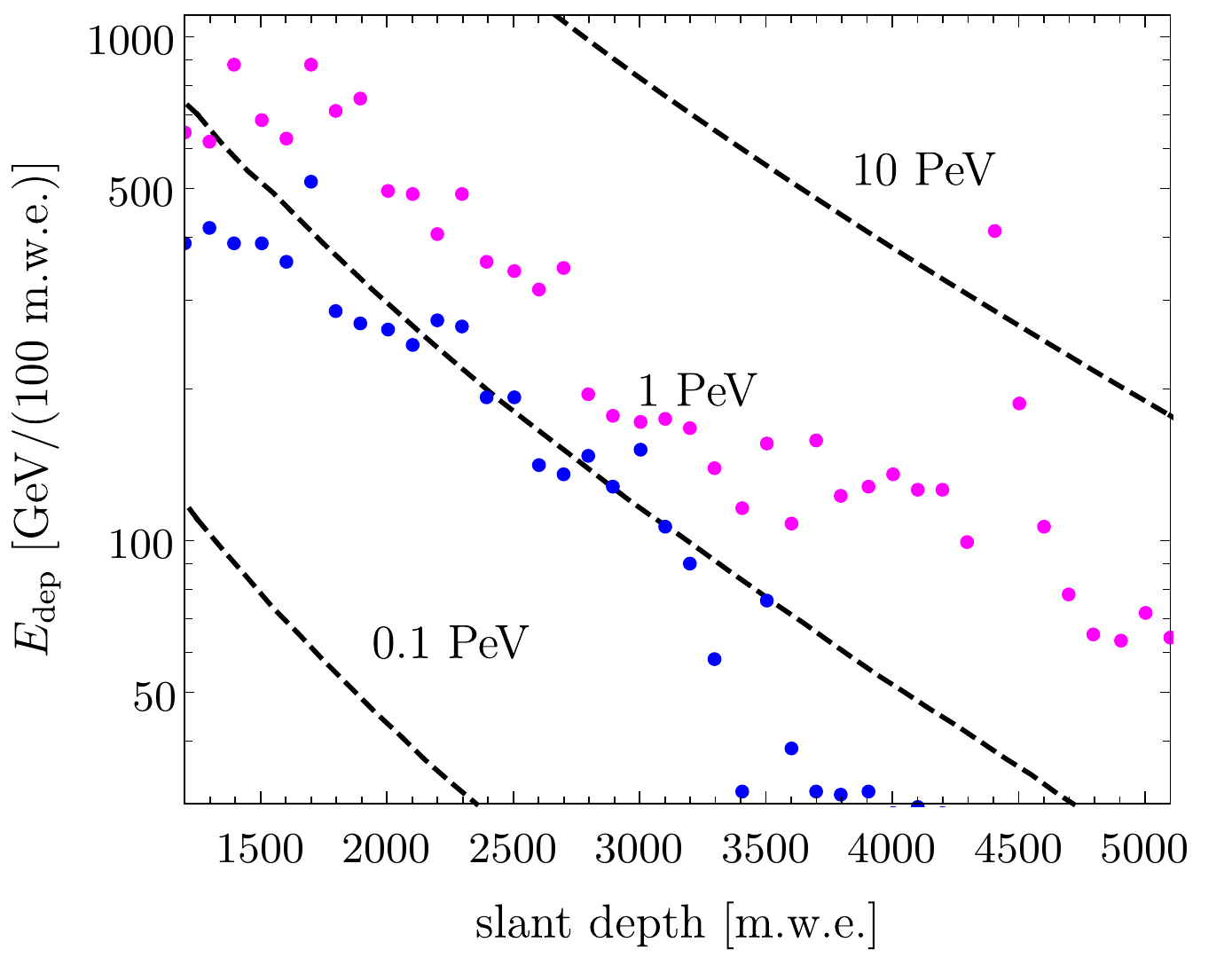}}
\caption{
{\bf Left.} Energy depositions 
in 100 meter intervals of ice or water for two muons 
of $E_\mu=0.5,1,3,5,10$ TeV and for two bundles from 1 PeV proton showers;
the total energy deposition in the interval is proportional to the volume of each blob, 
whereas the shaded region represents the IceCube volume: 1 km$^3$ at depths between 2450 and 1450 m. 
{\bf Right.} Average (over $10^4$ showers)
energy depositions in 100 meter intervals for bundles from proton primaries of different energy (we have included
the two PeV bundles on the left figure also in this plot).
\label{fig4}}
\end{center}
\end{figure}
In Fig.~\ref{fig4} we provide examples of the propagation 
of muons and of muon bundles through several km of ice, together
with the average energy deposited per 100 meters at different depths for bundles from proton 
primaries of $10^5$, $10^6$ and $10^7$~GeV. The average in the plot 
is obtained for $10^4$ showers of each energy;
it approximately 
scales like $E^{0.81}$ with the shower energy and like $\exp\left(-{X\over 1100 {\rm \; m.w.e.}}\right)$ 
with the slant depth.

\section{Neutrino events within a bundle\label{sec4}}

Our objective is to establish criteria to separate muon bundles that include a 
neutrino interaction from those bundles that do not. These criteria or {\it cuts} 
should be very efficient eliminating plain
bundles while selecting a significant fraction of the events with a neutrino interaction. 

For each event we define four basic parameters related to observables that can be measured 
with different degree of precision in telescopes like IceCube or KM3NeT:

\begin{enumerate}

\item $X_A$: age of the track, {\it i.e.}, the slant depth from the ground to the point of entry in the
detector. $X_A$ depends on the  inclination and the coordinates of the event.

\item $E_{\rm max}$: maximum energy deposition within a 100 meter interval along the track crossing
the detector. 

\item $E_{-}$: total energy deposited in the detector {\it before} the maximum deposition 
$E_{\rm max}$ divided by the number of 100 meter intervals. Our unit length is set at 100 meters,
the typical separation between strings at km$^3$ telescopes.

\item $E_{+}$: total energy in the detector after 
$E_{\rm max}$ divided by the number of 100 meter intervals.

\end{enumerate}
The number of 100 meter intervals before and after the maximum  
deposition will depend, like $X_A$, on the inclination and coordinates of each event. 
We will define cuts in terms of the ratios $\Em/E_-$ and $E_+/E_-$.

Let us first consider charged current (CC) $\nu_e$ events, with all the neutrino energy deposited 
in a single 100 m interval.
A typical 1 TeV event will come together with the weak muon bundle of a $E\le 20$ TeV shower, 
able to reach the telescope only from vertical directions. These events would imply a value of 
$\Em \ge 30 E_-$. A similar deposition $\Em$ could as well be produced by a muon that 
reaches the telescope with an energy of, for example, $2$~TeV. However
{\it (i)} such muons deposit around 50 GeV in each 100 m interval previous to 
$\Em$, and {\it (ii)} they usually appear inside
more energetic showers, together with other muons that also contribute to $E_-$ and reduce 
the value of $\Em/E_-$. In addition, this type of depositions subtracts a significant fraction of energy to the
muon, implying a drop in the signal after $\Em$. Notice that this effect would be absent when the energy
deposition is caused by the neutrino. Requiring that $E_+\ge 0.9 E_-$
we would make sure that those events do not pass the cut.

If we increase the energy by a factor of 10 and target 10 TeV CC $\nu_e$ events, two competing 
effects are noticeable. On one hand,
stochastic energy depositions grow linearly with the energy of a muon, while the growth of its
continuous energy loss is a bit slower.\footnote{Notice that this includes ionization but also radiative
processes of $\nu<10^{-2.5}$.} This first effect suggests that we should increase the
minimum value of $\Em/E_-$ required to select  neutrino
events. On the other hand, however, the scaling also implies stronger 
muon bundles giving a more sustained and regular deposition: if a 50 TeV muon reaches the detector 
after crossing a depth $X_A$, it is
likely that other less energetic muons in the same bundle will reach as well. 
Although both effects tend
to cancel, we conclude that a more effective cut to select $\nu_e$ events must depend on $\Em$:
\beq
{\Em \over E_-} > 20 \left(1+ {2 \over {1+ \exp \left( 2 - {\Em \over 1\;{\rm TeV}}  \right)}} \right)
\,; \hspace{0,5cm} 0.9 \le {E_+ \over E_-} \le 1.5  \;\;{\rm or}\;\; {E_+ \over E_-} \le 0.02\,.
\label{nue}
\eeq
As we have already mentioned, 
the cut is a condition on the ratio between the energy deposited after
and before the maximum deposition. 
If $E_+/E_- < 0.9$ then there is a chance
that $\Em$ has been caused by a very energetic muon, whereas a {\it revival} of the signal by a
factor of 1.5 after $\Em$, $E_+/E_-  > 1.5$, is only expected in $\nu_\mu$ events (see below). 
The $E_+<0.02 E_-$ possibility in the search for $\nu_e$ events 
is added to include neutrinos interacting {\it after} all muons in the bundle have stopped. 
Finally, we will also require that the track intersecting the detector must have a minimum length of 
500 m, with at least 200 m before the maximum energy deposition ({\it i.e.}, $E_-$ is obtained as 
the average over at least two 100 meter intervals).

The characterization of CC $\nu_\mu$ events is equally simple. The two main differences with the case
just discussed are that 
{\it (i)} the $\nu_\mu$ will deposit in the interaction point 
only a  fraction of its energy and {\it (ii)} it will create a muon of similar
energy. Again, it is essential that the main contribution to the atmospheric neutrino flux comes from 
primaries of energy (per nucleon) just 5--20 times larger. A typical event will consist of a
$\nu_\mu$ together with a leading muon and a bundle: first
 the propagation (a large enough age $X_A$) weakens the muon track entering the
detector, then there is a significant energy deposition ($\Em \gg E_-$)
followed by a track that is revived by the final muon ($E_+>E_-$). We can use
\beq
{\Em \over E_-} > 20 \left(1+ {2 \over {1+ \exp \left( 2 - {\Em \over 1\;{\rm TeV}}  \right)}} \right)
\,; \hspace{0,5cm} {E_+ \over E_-} > 1.5\,.
\label{numu}
\eeq
This condition is fully effective when the track inside the detector includes at least two 100 m length 
intervals before and after $\Em$. 

\begin{figure}[t]
\begin{center}
\includegraphics[width=0.49\linewidth]{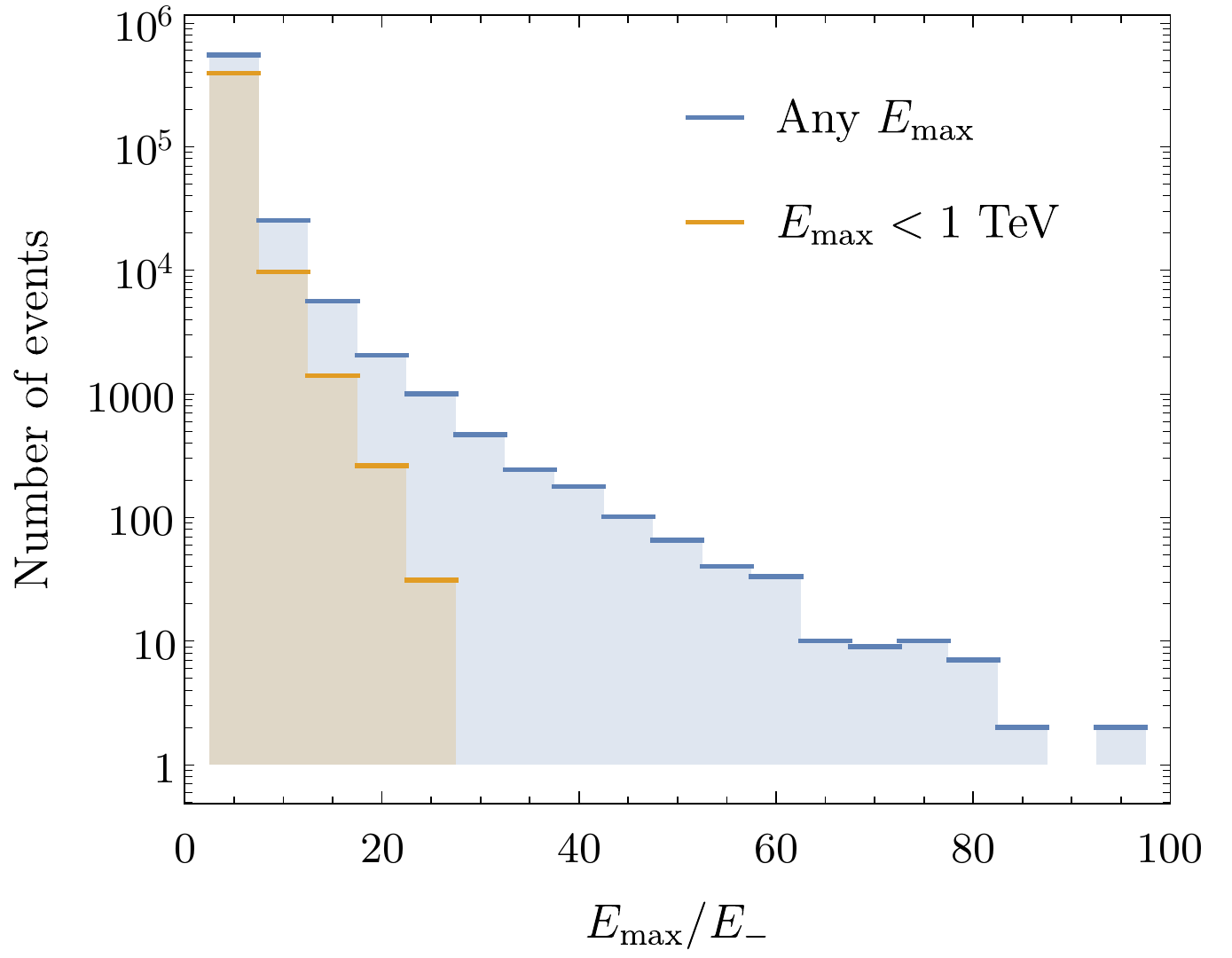}\hfill
\includegraphics[width=0.49\linewidth]{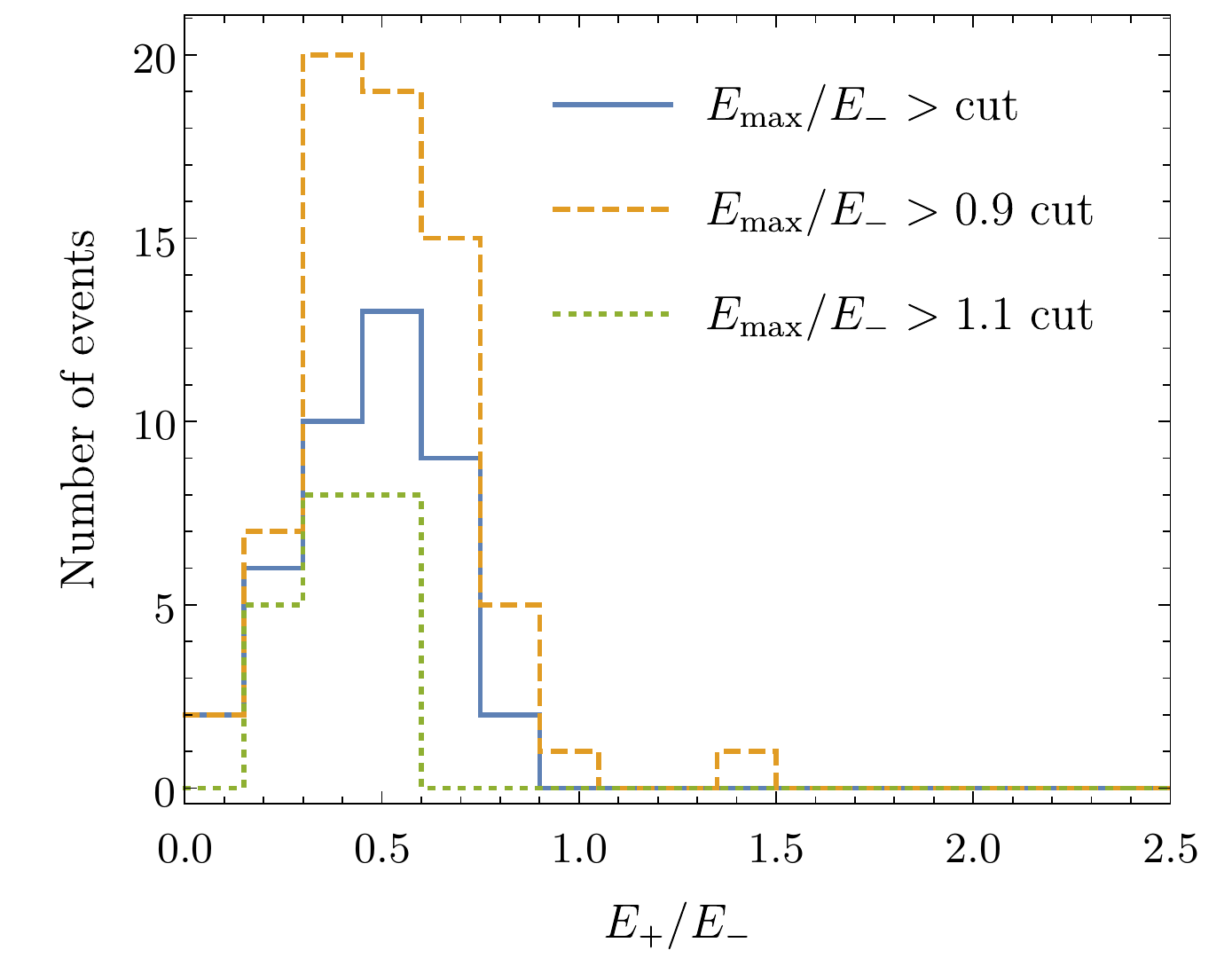}
\caption{
{\bf Left.} Distribution of $\Em/E_-$ for $6\times 10^5$ bundle ``events'' (see text)
from $10^4$ proton showers of $10$ PeV.
We have separated the events with $\Em<1$ TeV.
{\bf Right.} Distribution of $E_+/E_-$ for the 42 events that pass the first cut. We include the distribution when
 we reduce (dashes) or increase (dots) in a 10\% the $\Em/E_-$ cut.
\label{fig5}}
\end{center}
\end{figure}
We have simulated and analyzed a sample of $10^4$ muon bundles from proton and He showers of 
energy between 
$10^4$ and $10^8$~GeV at $X_A> 1500$ m.w.e. Our procedure has been the following. 
First we generate the bundle. Then we make a Monte Carlo simulation of its propagation 
through the ice or water, where it defines a {\it track} of energy depositions. We take the track after a 
slant depth of 1500 m.w.e. and divide it in 500 or 900 meter intervals (``short''  and  ``long''  events): 
each one of these intervals might be intersecting the telescope and define an event. For each event, we 
take 100 meter segments, we determine the total energy deposition in each segment, we find $\Em$, 
$E_-$ and $E_+$ and finally we apply the cuts. 

We find no segment with 5 or 9 length intervals ({\it i.e.}, a
500--900 meter track inside the detector) that passes the cuts established above and gives a false positive.
In Fig.~\ref{fig5} (left) we plot the distribution of $\Em/E_-$ for $E_{\rm shower}=10$ PeV, which is
the energy with the largest fraction of events passing the first cut (all energies give qualitatively similar results). 
In the plot we separate the
events with $\Em\le 1$ TeV, which have the cut at $\Em/E_-\approx 30$ (at higher values of $\Em$ the cut is
closer to $60$). When we apply this first cut plus the requirement of at least two length intervals before
$\Em$, 42 out of the $6\times 10^6$ events survive. Then we apply the cut on $E_+/E_-$. In Fig.~\ref{fig5} (right)
we show how this variable is distributed among the 42 events: none of them passes the second 
requirement to be classified as a $\nu_e$ ($0.8< E_+/E_-< 1.5$) or a $\nu_\mu$ ($1.5< E_+/E_-$) event. 

If we relaxed in a 10\% the cut on $\Em/E_-$, 71 events would pass it 
and 2 of them would give a false positive (they both would be declared $\nu_e$ events, see Fig.~\ref{fig5}). 
In contrast, a 10\% increase in the minimum value of $\Em/E_-$ implies that only 21 events pass the first cut
and all of them are clearly
excluded by their value of $E_+/E_-$.
The events more likely to give a false positive appear when a
single muon is produced with a large fraction (above 1\%) of the shower energy \cite{Gamez:2019dex}. 

As for the real neutrino events, when we include an arbitrary neutrino interaction that passes the cuts 
we find that the prescription separating $\nu_\mu$ from $\nu_e$ events is very efficient. In particular, 
we find that $\nu_\mu$ CC interactions are never taken as a $\nu_e$ event, whereas the 
opposite case ($\nu_e$ interactions confused with a $\nu_\mu$  CC event) has a frequency below 10\%.

\begin{figure}[t]
\begin{center}
\includegraphics[width=0.49\linewidth]{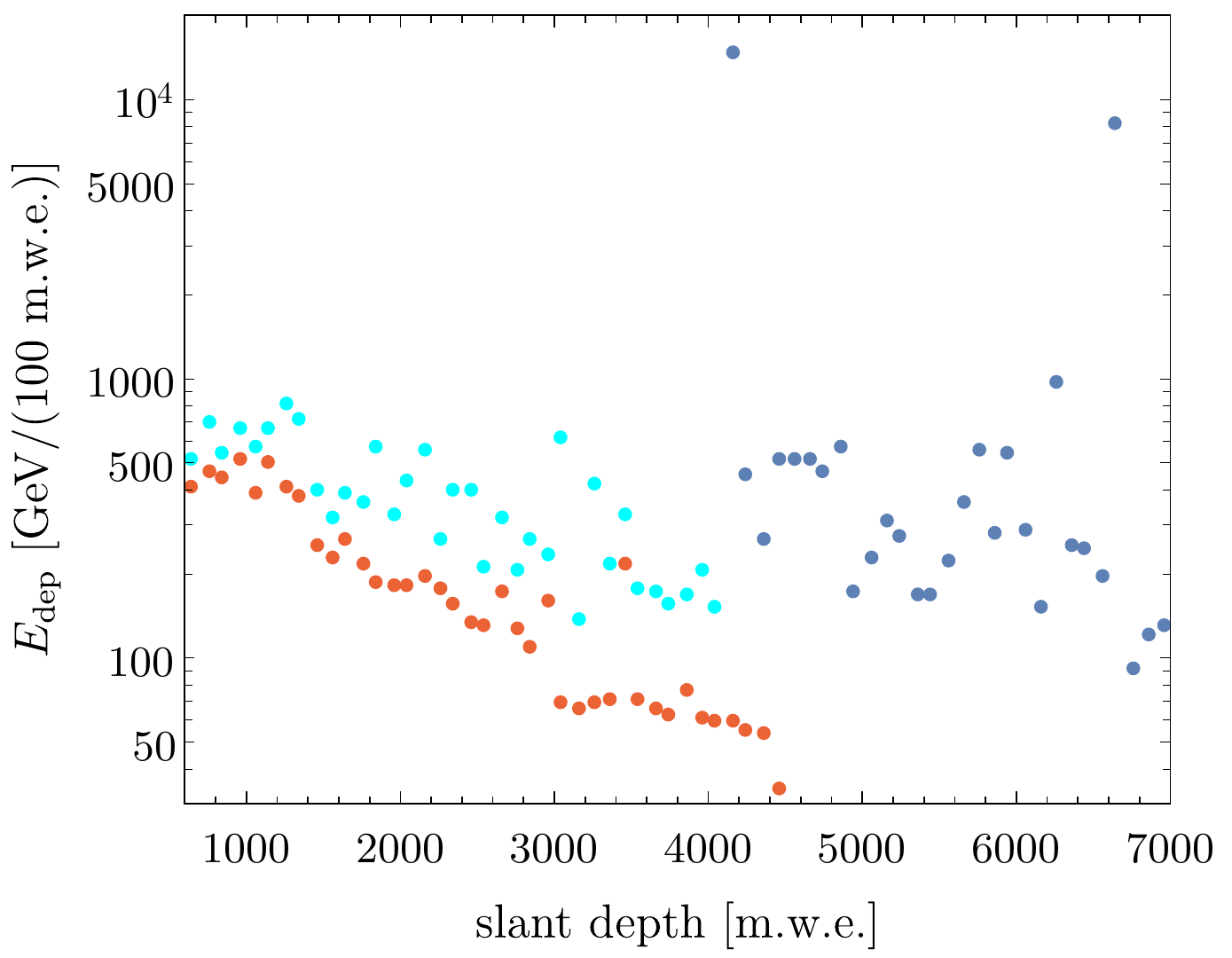}\hfill
\includegraphics[width=0.49\linewidth]{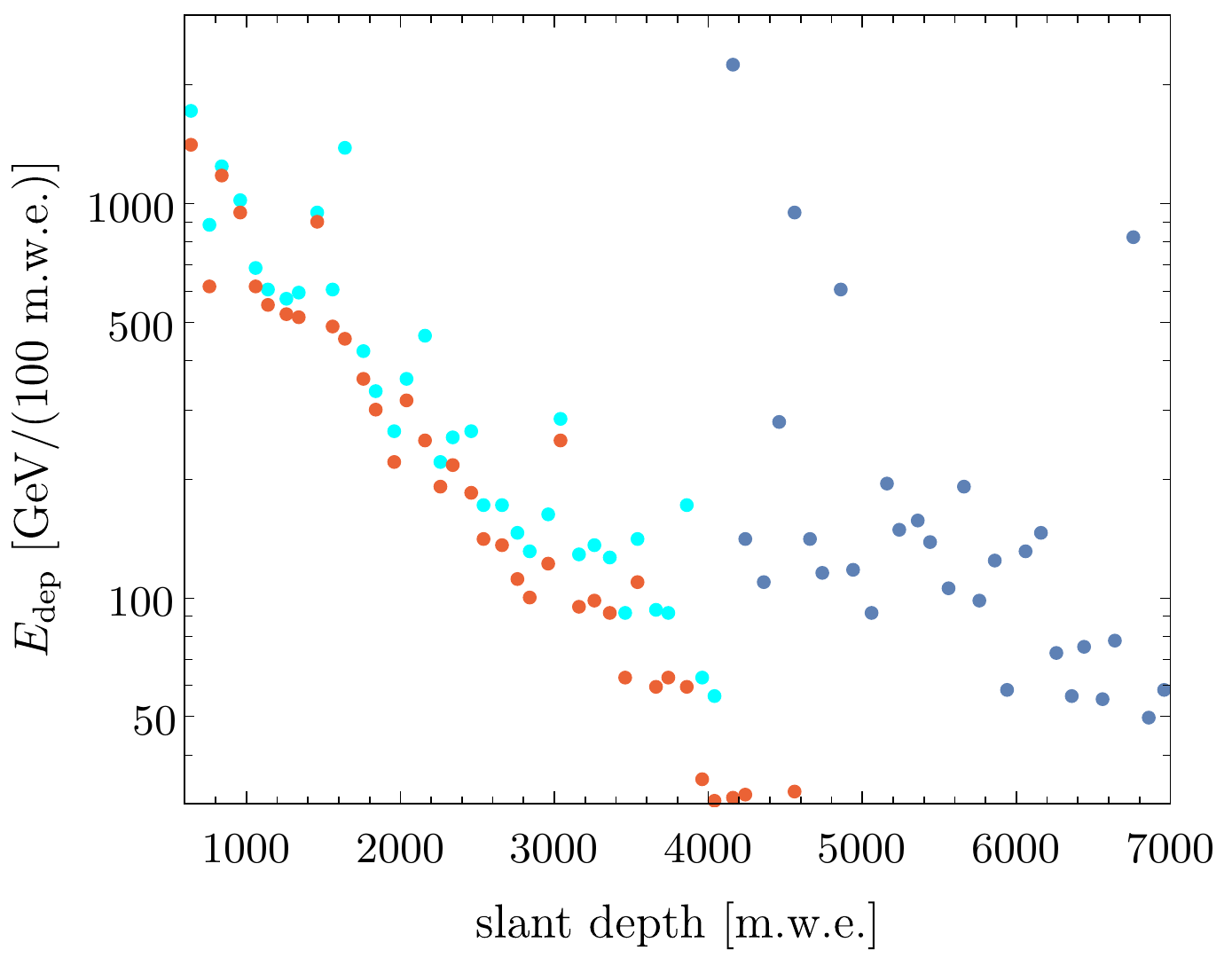}
\caption{
Energy depositions in 100 meter intervals of ice or water for 
a $10^6$ GeV proton (left) or a He (right) shower. Cyan and blue dots indicate, respectively,
the depositions before and after the $\nu_\mu$ CC interaction (see text). We include in red dots the
depositions of the muon bundle without the leading muon; the $\nu_e$ energy deposition defining $\Em$
in this case has been omitted.
\label{fig6}}
\end{center}
\end{figure}

Let us illustrate these results with a couple of examples. 
In Fig.~\ref{fig6} we provide the energy depositions produced by a $10^6$ GeV
proton (left) or a He (right) shower. The
red dots correspond to a typical muon 
bundle for a $\nu_e$ event (with no leading muon);  the proton shower in the plot
includes  10 muons of energy between 500 GeV and 2.1 TeV, whereas the He shower 
generates 17 muons of energy between 500 GeV and 3.6 TeV. In both cases, we see 
that any $\nu_e$
CC event of energy above 10 TeV (omitted in the plot) at a slant
depth $X_A\ge 2000$ m.w.e.~would pass the cut for $\nu_e$ events defined in 
Eq.~(\ref{nue}). 
In a contained $\nu_\mu$ event (cyan and blue dots in the same figure) we expect a 
leading muon after the interaction: we have added  to the muon bundle
a 30 TeV leading $\nu_\mu$ plus a 14 TeV muon (proton shower) or 
a 14 TeV $\nu_\mu$ plus a 10 TeV muon (He shower). The $\nu_\mu$ experiences then a CC 
interaction of inelasticity 0.46 and 0.22, respectively, in both cases at $4100$ m.w.e. If the
interaction occurred inside the detector, the revival of the track produced by the final 
muon would imply that the event passes the $\nu_\mu$ cut. If 
the same deposition $\Em$ 
were produced by a an isolated 10 TeV muon, instead of stronger
the track afterwards would have become significantly weaker.

\section{Summary and discussion\label{sec5}}

The determination of the atmospheric neutrino flux at energies from about 1 TeV up to several 100 TeV 
is essential both in the search for atmospheric 
charm and for a precise characterization of the high energy diffuse flux recently discovered by IceCube. 
However, this atmospheric flux is difficult to access with $\nu$ telescopes, as at $E\approx 10$ TeV 
it seems to be 5--10 times weaker than the astrophysical one. Any possibility to disentangle 
these two components in the flux and search for neutrinos from charm decays  
requires the detection of $\nu$ interactions 
in down-going events, where the presence of additional muons will reveal the atmospheric origin.

Here we have explored that type of events. Our analysis focuses on the muon bundle produced 
in the core of the air shower together with 
the neutrino. In particular, we have studied the energy depositions as the
bundle propagates in ice or water. The longitudinal pattern of depositions would translate into a 
particular signal in a km$^3$ telescope. Our objective has been to show that this pattern could
be different enough when it includes a neutrino interaction.

Our first observation has been that most atmospheric neutrinos are produced inside air showers that
are just ten times more energetic. As a consequence, its relative effect  on the signal 
associated to the muon bundle tends to be very large. The typical topology is a weak signal entering
the detector, followed by a large energy deposition, and finally a stronger signal in case of a CC $\nu_\mu$ 
interaction or a weak one in a $\nu_e$ or NC interaction. Generically,  neutrino events imply a signal
that increases with the slant depth inside the telescope, while muon bundles tend to imply the opposite 
effect.

We have defined cuts based on the ratios $\Em/E_-$ and $\Em/E_+$ (see Section~\ref{sec4})  that seem to
exclude muon bundles of any energy. A muon can certainly 
have a stochastic deposition of half its energy, but not
without leaving a trace both before ($E_-$) and after ($E_+$) this $\Em$.
In $10^4$ simulations of muon bundles, we find that 
when the ratio $\Em/E_-$ is very large then the signal $E_+$ is significantly 
weaker (relative to $E_-$)
than in a CC $\nu_\mu$ or a $\nu_e$ event  ($E_+ < 0.8\,E_-$). The bundle events that are 
closest to the cuts include one single muon carrying a significant fraction of the shower energy
that deposits a large fraction of its energy when the rest of the bundle is 
already weak. Actually, the search for this type of muon events could be of interest by 
itself \cite{IceCube:2015wro,Fuchs:2017nuo,Gamez:2019dex} and seems also possible.

Our results should be considered just a first step in the search for neutrino interactions in 
down-going events at $\nu$ telescopes. 
We show that there are basic physics criteria that could separate these 
events from plain muon bundles. Of course, to determine in detail whether or not an analysis along these
lines could give positive results in actual observations would depend on the experimental conditions
(volume, energy resolution, triggers, etc.) at each observatory.
Neutrino telescopes have been built to look for high energy sources and avoid the atmospheric
background. However, they have also pursued other more {\it unlikely} but equally interesting 
objectives: IceCube 
has been able to define a strategy
to target transient event of energy as low as 1--10 GeV \cite{IceCube:2021jwt}, to look for high $p_T$ muons
\cite{Abbasi:2012kza,Soldin:2018vak}, to determine the atmospheric muon flux at $E_\mu\ge 10$ TeV 
\cite{IceCube:2015wro}
or to reconstruct starting muon tracks \cite{IceCube:2021oqo}. A more precise
characterization of the atmospheric neutrino flux at 1--100 TeV seems a very interesting objective
as well.

\section*{Acknowledgments}

This work was partially supported by the Spanish Ministry of Science, Innovation and Universities
(PID2019-107844GB-C21/AEI/10.13039 /501100011033) and by the Junta de Andaluc{\'\i}a 
(FQM 101, SOMM17/6104/UGR, P18-FR-1962, P18-FR-5057). MGG acknowledges 
a grant from {\it Programa Operativo de Empleo Juvenil} (Junta de Andaluc\'\i a).
The work of GHT has been funded by  the program {\it Estancias Postdoctorales en el Extranjero 2019-2020}
of CONACYT, Mexico. GHT also acknowledges Prof.~Pablo Roig for partial support through 
{\it C\'atedra Marcos Moshinsky} (Fundaci\'on Marcos Moshinsky). 

\vfil\eject

\appendix

\section{Neutrino yields\label{appA}}

In our parametrization of the yields in proton showers 
we have separated the conventional $f_{p\nu}^{\rm conv}(x,E)$ and the prompt 
$f_{p\nu}^{\rm charm}(x,E)$ contributions. The yields refer to
the sum of $\nu_i+\bar \nu_i$, with $i=e,\mu$, and we express them in terms of four energy and flavor dependent 
parameters ($A$, $B$, $C$, $D$) as:
\beq
f_{p\nu}(x,E) = A\,x^{-B}\,e^{-Cx}\,(1-x)^D\left( 1+ \sqrt{m_\mu\over x E - m_\mu} \right)^{\!-4},
\eeq
where $m_\mu$ is the muon mass.
From the CORSIKA-SIBYLL 2.3C simulation ($10^4$ showers of each energy with $E_{\rm min}=10^{-3}E_{\rm shower}$ 
and 50 showers with $E_{\rm min}=1$ GeV) we deduce the value of the 4 parameters for each flavor at six different proton energies, and then we interpolate (linearly
in $\log E$) inside each energy interval.

\begin{table}
\begin{center}
\begin{tabular}{c|c c c c c c c |} 
 \hline
$E$ [GeV] & $10^3$ & $10^4$ & $10^5$ & $10^6$ & $10^7$ & $10^8$ \\  [0.3ex] 
 \hline\hline
$E\times A^{\rm conv}_{\nu_\mu}$ & 10.0 & 37.7 & 38.8 & 29.5 & 26.9 & 27.2 \\ 
 \hline
$B^{\rm conv}_{\nu_\mu}$ & 2.40 & 2.30 & 2.50 & 2.65 & 2.70 & 2.70 \\ 
 \hline
$C^{\rm conv}_{\nu_\mu}$ & 2.0 & 2.2 & 2.4 & 2.7 & 4.0 & 8.0 \\ 
 \hline
$D^{\rm conv}_{\nu_\mu}$ & 3.7 & 3.8 & 4.0 & 4.0 & 4.0 & 4.0 \\ 
 \hline
 \hline
$E\times A^{\rm conv}_{\nu_e}$ & 0.55 & 1.14 & 2.18 & 1.40 & 1.35 & 1.75 \\ 
 \hline
$B^{\rm conv}_{\nu_e}$ & 2.60 & 2.50 & 2.50 & 2.65 & 2.70 & 2.70 \\ 
 \hline
$C^{\rm conv}_{\nu_e}$ & 3.0 & 3.0 & 4.0 & 4.5 & 5.0 & 5.0 \\ 
 \hline
$D^{\rm conv}_{\nu_e}$ & 4.6 & 4.7 & 5.0 & 5.0 & 5.0 & 5.0 \\ 
 \hline
 \hline
\end{tabular}
\caption{Energy-dependent parameters defining $f_{p\nu_\mu}^{\rm conv}(x,E)$ and $f_{p\nu_e}^{\rm conv}(x,E)$
at $\theta_z=45^\circ$. 
\label{table1}}
\end{center}
\end{table}

\begin{figure}[t]
\begin{center}
\includegraphics[width=0.49\linewidth]{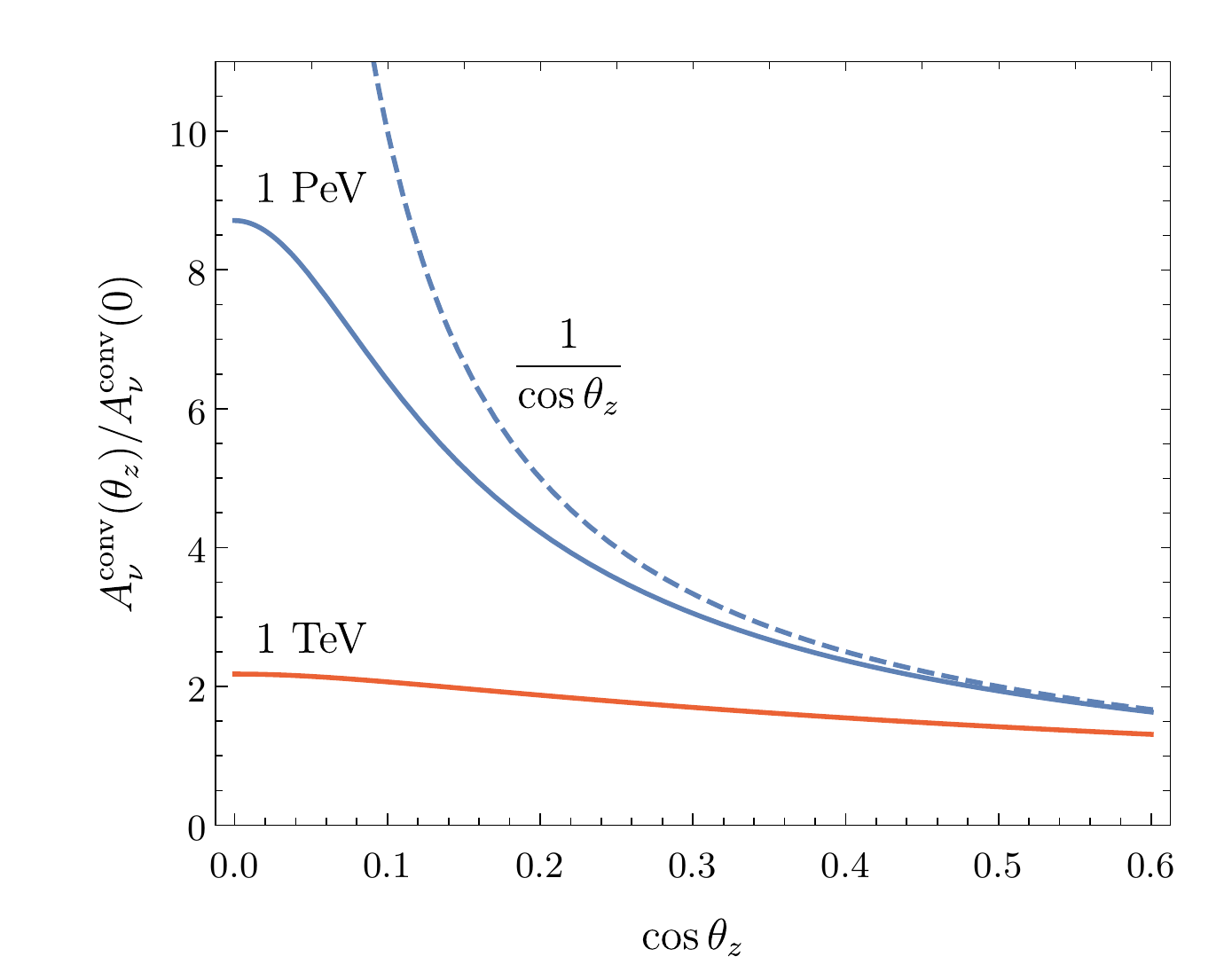}
\caption{Angular dependence for the normalization of the conventional $\nu_{e,\mu}$ yields for proton primaries at 
1 TeV and 1 PeV. We include in dashes $(\cos\theta_z)^{-1}$.
\label{fig7}}
\end{center}
\end{figure}

For the conventional yield at $\theta_z=45^\circ$ 
we obtain the values given in Table \ref{table1}. The angular dependence (see Fig.~\ref{fig7}) may be 
described in terms of the zenith angle of the line sight at 
$h=30$ km, $\theta^*(\theta_z)$, defined in \cite{Lipari:1993hd}:
\beq
\tan \theta^* = {R_\oplus \sin\theta_z\over \sqrt{R_\oplus^2 \cos^2\theta_z + \left( 2 R_\oplus + h \right) h }}\,,
\label{theta}
\eeq
with $R_\oplus$ the radius of the Earth. We fit
\beq
A^{\rm conv}_{\nu}(\theta_z) \approx A^{\rm conv}_{\nu}(45^\circ)\,
{\sqrt{300\;{\rm GeV}\over E} +\cos 44.73^\circ \over  \sqrt{300\;{\rm GeV}\over E} + \cos\theta^*}.
\eeq

For the neutrinos from charm 
decays we 
obtain similar $\nu_\mu$ and $\nu_e$ yields and no energy dependency in the 4 parameters:
\beq
A^{\rm charm}_{\nu_i}=1.0\times 10^{-4}\,;\;\;
B^{\rm charm}_{\nu_i}=1.8\,;\;\;
C^{\rm charm}_{\nu_i}=10.0\,;\;\;
D^{\rm charm}_{\nu_i}=5.0\,, 
\eeq
with $i=(\nu,\,e)$.

%%%%%%%%%%%%%%%%%%%%%%%%%%%%%%%%%%%%%%%%%%%%%%%%%%%%%%%%%%%%%%%%%%%%%%%%%%%%%%%

\vfil
\eject
%%%%%%%%%%%%%%%%%%%%%%%%

\end{document}